\documentstyle[epic,eepic]{article}
\begin{document}
\def\dr{\rangle\!\rangle}
\def\dl{\langle\!\langle}
\def\ad{a^{\dag}}\def\bd{b^{\dag}}\def\hU{\hat U}\def\hUd{\hat U^{\dag}}
\def\<{\langle}\def\>{\rangle}\def\hb{\hbar}\def\cd{c^{\dag}}
\def\hr{\hat\varrho}
\def\pni{\par\noindent}
\title{Equivalence between squeezed-state \\ 
and twin-beam communication channels}
\author{G. M. D'Ariano and M. F. Sacchi\\
Dipartimento di Fisica \lq\lq A. Volta\rq\rq, 
Universit\`a degli Studi di Pavia,\\
via Bassi 6, I-27100 Pavia, ITALY\\
e-mail: dariano@pv.infn.it\hspace{10pt} msacchi@pv.infn.it}
\date{}
\maketitle
\begin{abstract}
We show the equivalence between two different communication 
schemes that employ a couple of modes of the electromagnetic field. 
One scheme uses unconventional heterodyne 
detection, with correlated signal and image band modes in a twin-beam 
state from parametric downconversion. 
The other scheme is realized through a complex-number coding over 
quadr\-ature-squeezed states of two 
uncorrelated modes, each detected by ordinary homodyning. 
This equivalence concerns all the stages of the communication channel: 
the encoded state, the optimal amplifier for the channel, the master 
equation modeling the loss, and the output measurement scheme. 
The unitary transformation that connects the two communication schemes 
is realized by a frequency conversion device.
\end{abstract}
\section{Introduction}
The fundamental theorems of quantum communication theory 
\cite{hol,yoz} establish an upper bound on the mutual information for 
all possible coded inputs, and all kinds of output detection 
at the channel. 
These results, when applied to a single narrow-band linear bosonic
channel \cite{caves}, specify the ultimate quantum capacity per use 
(i.e. the maximum mutual information {\em per mode} 
subjected to any constraint 
on the channel), which, for fixed average-power constraint, could be ideally 
achieved by direct detection of number-states 
with thermal {\em a priori} input probability. 
Any other communication scheme can transmit just less information than 
the number-state (NS) channel. 
At high power levels, for example, the ideal NS channel achieves 
1.44 bits per use more than the ideal coherent-state 
(CS) channel, and 0.44 bits more than the quadrature-squeezed-state (QS) 
channel \cite{caves}.
(The CS channel utilizes heterodyne detection of CS, with Gaussian 
{\em a priori} probability; the QS channel uses homodyne detection of QS 
having Gaussian probability and optimal fraction of squeezing photons). 
\par Although the NS channel is the optimal one, 
several reasons relegate it to a purely theoretical plan, and 
lead to consider the QS 
channel as a more realistic (and not much less efficient) 
communication scheme. 
First of all, it is a challenge to achieve number eigenstates 
experimentally. 
Second, the optimal amplification for the QS communication channel 
is achieved by a phase-sensitive ideal amplifier, which can be 
realized experimentally, 
whereas a concrete realization of the ideal 
photon number amplifier for the NS channel is still unknown 
\cite{yy,gmd}. Finally, the detrimental effect of loss, widely neglected in 
the literature 
\cite{nota}, easily degrades the NS channel: 
for example, for ten photons of average power, a signal attenuation of 0.2dB 
is enough to degrade the NS channel under the capacity of the CS one, 
and for higher power the effect is 
more dramatic \cite{next}. 
The QS channel is more robust to losses than 
the NS channel, and at low powers and not too high losses it remains 
above the capacity level of the CS channel \cite{next}. 
\par In this paper we show the equivalence between two communication 
schemes that employ complex-number encoding carried by two-mode states. 
The potentiality of a quantum communication channel that uses 
two modes has already been noticed, in particular for 
phase-modulation-based digital 
communications \cite{shap}. The two equivalent schemes are the 
following: \begin{description}
\item[\em (i)] a complex number is encoded on a 
twin-beam state generated by parametric downconversion; decoding 
is achieved through unconventional heterodyne 
detection of both the signal and image-band modes which form the 
correlated twin-beam; 
\item[\em (ii)] the real and the imaginary part 
of the complex number are independently encoded over two 
uncorrelated QS pertaining two different modes; the two QS are decoded 
through ordinary homodyne detection on each mode separately.
\end{description}
\par The equivalence between the above schemes 
involves all the stages of the communication channel: 
coding, decoding and optimal amplification. Moreover, it includes 
the master equation that models the loss along the line. 
The equivalence is realized by a unitary transformation that 
physically corresponds to 50-50 frequency conversion. 
The equivalence between the two channels is relevant for applications, 
because the heterodyne twin-beam 
communication channel is easier to achieve experimentally 
than the QS channel. 
\par The paper is organized as follows. In Sec.~II we describe the 
twin-beam communication scheme. We introduce the eigenstates of the 
heterodyne detector photocurrent, and give a scheme that realizes 
states that approach such eigenstates used to encode the 
transmitted information. 
In Sec.~III we discuss the two-mode master equation that models 
the effect of both losses and linear phase-insensitive amplification, 
and then we derive a Fokker-Planck equation for the unconventional-heterodyne 
probability. 
In Sec.~IV the channel equivalence is analyzed. 
We show that the 50-50 frequency conversion disentangles a twin-beam state 
into a couple of QS, whereas, at the same time, it changes 
a phase-insensitive amplifier (PIA) into two independent 
phase-sensitive amplifiers (PSA). 
On the other hand, the master equation for the loss 
turns out to be invariant under frequency conversion. 
In Sec.~V the Fokker-Planck equations for the two schemes are compared, 
and the mutual information that characterizes 
the communication channels is evaluated. 
Finally, maximization over input 
probabilities is carried out to get the capacity per use. 
\section{The twin-beam communication scheme}
\subsection{Unconventional-heterodyne detection}
Under ideal conditions, the output photocurrent of a heterodyne 
detector corresponds \cite{yush,shwa} to the 
complex operator $\hat Z=a+\bd$, $a$ and $b$ denoting the 
annihilator of the signal and the image-band mode, respectively. 
Ordinary heterodyne detection corresponds to the joint measurement of both 
quadratures of the field $a$, with $b$ as the vacuum image-band mode 
responsible of the 3dB heterodyne added noise. 
The communication scheme here presented is 
based on unconventional heterodyne detection, namely  with the signal and 
image-band modes both nonvacuum. The eigenvector of $\hat Z$ with 
complex eigenvalue $z$ can be written in several equivalent forms. In 
terms of the eigenvectors $|x\rangle_{\phi}$ of the quadrature 
$\hat{X}_{\phi}=\frac{1}{2}(c^{\dag}e^{i\phi}+\hbox{h.c.})$ of 
the mode $c=a,b$ one has \cite{shwa,rc}
\begin{eqnarray}
|z\rangle\!\rangle &=& \int_{-\infty}^{+\infty}\frac{dx}{\sqrt{\pi}}
e^{2ix\hbox{\scriptsize Im}z}|x\rangle_{0}\otimes|\hbox{Re}z-x
\rangle_{0}\nonumber  \\
&=&\int_{-\infty}^{+\infty}\frac{dy}{\sqrt{\pi}}
e^{-2iy\hbox{\scriptsize Re}z}|y+\hbox{Im}z\rangle_{\pi/2}
\otimes |y\rangle_{\pi /2}\;,\label{zeta1}
\end{eqnarray}
where $|\psi \rangle \otimes |\varphi \rangle $ denotes a vector 
in the two-mode Hilbert space ${\cal H}={\cal H}_{a}\otimes{\cal H}_{b}$ 
[the notation $|\ \rangle\!\rangle$ remembers 
that the state is a two-mode one]. 
A number representation of the state (\ref{zeta1}) can be found in Ref. 
\cite{rc}. The eigenstate for zero 
eigenvalue reads
\begin{eqnarray}
|0\rangle\!\rangle=
{1\over\sqrt\pi}
\exp(-\ad\bd)|0\>\otimes|0\>=
\frac{1}{\sqrt{\pi}}\sum_{n=0}^{\infty}(-)^{n}
|n\rangle\otimes|n\rangle\;.\label{twin}                              
\end{eqnarray}
It is convenient to write the eigenstate of $\hat Z$ corresponding 
to complex eigenvalue $z$ as follows
\begin{eqnarray}
|z\dr={e^{|z|^2\over 2}\over\sqrt\pi}
\exp(-\ad\bd)|z\>\otimes|\bar z\>\;,\label{zeta}
\end{eqnarray}
where $|z\rangle $ denotes a customary coherent state. 
The states $\{|z\rangle\!\rangle\}$ are Dirac-normalized 
as $\langle\!\langle z|z'\rangle\!\rangle =\delta^{(2)}(z-z')$, 
and form a complete orthogonal set for $\cal H$. 
\par In the following, we will introduce a set of physical 
(normalizable) two-mode states that approach the (infinite-energy) 
eigenstates of $\hat Z$.
\subsection{Limited power near-eigenstates of the photocurrent}
The physical realization of states that approach the eigenstates 
$|z\dr$ in Eq. (\ref{zeta}) is suggested by recognizing that 
the zero-eigenvalue state (\ref{twin}) is just a  ``twin-beam'' 
at the output of a PIA in the limit of infinite gain \cite{mauro1}. 
All the other states with $z\neq 0$ can be approached by suitably displacing 
either a single mode $a$ or $b$, 
or both of them, and this can be done 
by means of high-transmissivity beam-splitters 
with a strong coherent local oscillator. Such physical realizations 
$|z\dr _{\lambda}$ can be written as follows
\begin{eqnarray}
|z\dr _{\lambda}=D_a(v)D_b(\bar w)e^{\hbox{{\scriptsize tanh}}^{-1}\,\lambda\,
(ab-a^{\dag}b^{\dag})\,}|0\>\otimes|0\>\,\quad (z=v+w)\;,\label{zetal}
\end{eqnarray}
where $D_c(u)=e^{uc^{\dag}-\bar uc}$ denotes the displacement operator 
of the mode $c=a,b$, and $e^{\hbox{{\scriptsize tanh}}^{-1}\,\lambda\,
(ab-a^{\dag}b^{\dag})\,}$ is the unitary transformation 
achieved by a PIA with gain $G=(1-\lambda ^2)^{-1}$. [The PIA Hamiltonian 
$(\ad\bd + \hbox {h.c.})$ corresponds to a ${\chi }^{(2)}$ parametric 
downconversion in the rotating-wave approximation and for classical 
undepleted pump]. 
The average number $N$ of photons {\em per mode} of the state (\ref{zetal}) 
is given by
\begin{eqnarray}
N={1\over 2}\<\ad a+\bd b\>={\lambda ^2\over 1-\lambda 
^2}+{|v|^2\over 2}+{|w|^2\over 2}\;.\label{mean}
\end{eqnarray}
The probability density of the photocurrent in the state 
$|u\dr _{\lambda}$ has the Gaussian form
\begin{eqnarray}
|\langle\!\langle z|u\rangle\!\rangle_{\lambda}|^{2}  
=\frac{1}{\pi\Delta_{\lambda}^{2}}\exp\left( {-\frac{|z-u|^{2}}
{\Delta_{\lambda}^{2}}}\right)    \;,\label{prob}
\end{eqnarray}
with variance 
\begin{eqnarray}
\Delta_{\lambda}^{2}=(1-\lambda )/(1+\lambda )
\;.\label{dellam}
\end{eqnarray}
\par\noindent Using the identity
\begin{eqnarray}
e^{-\xi(ab-a^{\dag}b^{\dag})}\,D_a(v)\,D_b(\bar w)\,
e^{\xi(ab-a^{\dag}b^{\dag})}=D_a(v\,\hbox{ch}\xi+w\,\hbox{sh}\xi)\,
D_b(\bar v\,\hbox{sh}\xi+\bar w\,\hbox{ch}\xi)\; 
\end{eqnarray}
Eq. (\ref{zetal}) rewrites as follows 
\begin{eqnarray}
|z\dr _{\lambda}=e^{\hbox{{\scriptsize tanh}}^{-1}\,\lambda\,
(ab-a^{\dag}b^{\dag})\,}\left|{v+\lambda w\over \sqrt{1-\lambda ^2}}
\right\>
\otimes\left|{\lambda \bar v+\bar w\over \sqrt{1-\lambda ^2}}
\right\>\,
\quad (z=v+w)\;.\label{down}
\end{eqnarray}
Hence, the state $|z\dr _{\lambda}$ can be also obtained 
through phase-insensitive amplification of input signal and idler 
which have the precise amplitude relation in Eq. (\ref{down}). 
Notice that the probability density 
(\ref{prob}) does not depend explicitly on $v$ and $w$ (the share of 
displacement of the two modes), while the mean number of photons does. 
The constraint $v+w=z$ over states (\ref{zetal}) and (\ref{down}) 
implies that for each $z$ there is a family of states 
approaching the eigenstate $|z\dr$ according to Eq. (\ref{prob}). 
For $v=w=z/2$, the most symmetrical state
\begin{eqnarray}
|z\dr _{\lambda}=e^{\hbox{{\scriptsize tanh}}^{-1}\,\lambda\,
(ab-a^{\dag}b^{\dag})\,}|z/2\Delta_{\lambda}\>
\otimes|\bar z/2\Delta_{\lambda}\>
\;\label{sim}
\end{eqnarray}
achieves the best phase sensitivity, as shown in Ref. \cite{rc}. 
For high photon number, the marginal probability density for the phase 
$\hat{\phi}=\arg(\hat{Z})$ is Gaussian, and in the limit of infinite PIA 
gain ($\lambda\rightarrow 1^-$) the r.m.s. phase sensitivity 
is optimized versus the total photon number $\bar n\equiv 2N$ 
to the value \cite{rc2}
\begin{eqnarray}
\delta\phi=\langle\Delta\phi ^2\rangle^{1/2}
\simeq{1\over\sqrt 2\,\bar n}\;.\label{1sun}
\end{eqnarray}
For a repeatable phase measurement scheme on a two-mode 
field, see Ref. \cite{proc}.
\par The couple of coherent states on the right of Eq. (\ref{sim}) 
can be generated by means of a single coherent state and a frequency 
conversion device with suitable pump strength and phase. 
Indeed, for any complex $\alpha $, one has
\begin{eqnarray}
\exp\left[ {\pi\over 4}\left( 
e^{i\arg \alpha}\ad b -e^{-i\arg \alpha}a\bd\right) \right] 
|\sqrt 2\alpha \>\otimes|0\>=|\alpha \>\otimes|\bar\alpha \>\;,
\label{11}
\end{eqnarray}
and the unitary operator on the left side of Eq. (\ref{11}) describes 
a frequency conversion device 
[in the parametric approximation of classical undepleted 
pump, this can be realized through a three-wave (or degenerate four-wave) 
mixing in a nonlinear $\chi^{(2)}$ ($\chi^{(3)}$) medium]. 
The two equivalent experimental set-ups 
to generate the state (\ref{sim}) are sketched in Fig.~1. 
\section{Losses and distributed amplification}
The effect of losses on the communication channel can be modeled by 
the following master equation
\begin{eqnarray}
\partial _t\hr={\cal L}_{\Gamma}\hr 
\doteq\Gamma\left\{(n_a+1)L[a]+(n_b+1)L[b]+
n_a L[\ad]+n_b L[\bd]\right\}\hr\;.\label{loss}
\end{eqnarray}   
In Eq. (\ref{loss}), the superoperator ${\cal L}_{\Gamma}$ 
gives the time derivative of 
the density matrix $\hr$ of the radiation state in the 
interaction picture, and acts on $\hr$ through the Lindblad superoperators 
$L[c]\hr=c\hr c^{\dag}-{1\over 2}(c^{\dag}c\hr+\hr c^{\dag}c)$ \cite{lind}. 
The damping rate $\Gamma$ is supposed to be equal for both modes, 
whereas the mean number of thermal photons $n_a$ and $n_b$ 
at the frequency of modes $a$ and $b$ can be neglected 
at optical frequencies. 
The absence of cross-terms that correlate the two modes is a 
consequence of the rotating-wave approximation assumed in the 
ordinary derivation of the master equation \cite{gard}. In a similar fashion, 
an active medium amplifier in the linear regime can be described by 
the superoperator
\begin{eqnarray}
{\cal L}_{\Lambda}=\Lambda\left\{(m_a+1)L[\ad]+(m_b+1)L[\bd]+
m_a L[a]+m_b L[b]\right\}
\;,\label{gain}
\end{eqnarray}   
where $\Lambda $ denotes the gain per unit 
time---i. e. the amplifier length---and 
$m_a$ and $m_b$ are related 
to the population inversion of the lasing levels at resonance with $a$ 
and $b$, respectively. Finally, a distributed parametric amplification 
with modes $a$ and $b$ correlated by a classical pump is 
represented by the following commutator
\begin{eqnarray}
{\cal L}_{K}=K[\ad\bd-ab,\cdot ]
\;,\label{comm}
\end{eqnarray}   
where $K$ is the gain per unit time and is related to the intensity 
of the pump.
\pni The following differential representation of modes
\begin{eqnarray}
a|z\dr=\left( {z\over 2}-\partial_{\bar z}\right)|z\dr \qquad\qquad  
\ad|z\dr=\left( {\bar z\over 2}+\partial_{z}\right)|z\dr \nonumber \\  
b|z\dr=\left( {\bar z\over 2}-\partial_{z}\right)|z\dr\qquad\qquad  
\bd|z\dr=\left( {z\over 2}+\partial_{\bar z}\right)|z\dr
\;\label{diff}
\end{eqnarray}
converts a two-mode master equation of the general form
\begin{eqnarray}
{\cal L}\hr _t&=&2\left\{(A+C_a)L[\ad]+(A+C_b)L[\bd]+
(B+C_a)L[a]+(B+C_b)L[b]\right\}\hr _t\nonumber\\
&+&K[\ad\bd-ab,\hr _t]
\;\label{me2}
\end{eqnarray}   
into the following Fokker-Planck equation 
\begin{eqnarray}
\partial _t P(z,\bar z;t)=
\left\{Q\left(\partial_z z+\partial_{\bar z}\bar z \right)
+2D\partial^{2}_{z\bar z}   \right\}
P(z,\bar z;t)\;,\label{fp2}
\end{eqnarray}
where $P(z,\bar z;t)\equiv\dl z|\hr _t|z\dr$ denotes the 
(unconventional) heterodyne 
probability density, and the drift and diffusion terms are given by 
$Q=B-A-K$ and $D=A+B+C_a+C_b$. 
Notice that the coefficients in the master equation (\ref{me2}) are not 
independent (the difference of the first two coefficients equals the 
difference of the last two in the curly brackets), 
and this is the condition under which all the 
derivatives in Eq. (\ref{diff}) gather to 
give the simple Fokker-Planck equation (\ref{fp2}). 
Hence, when considering the 
effect of loss in Eq. (\ref{loss}), the assumption of equal damping for 
the two modes is crucial. Of course, the same 
argument holds true for the gain $\Lambda$ in Eq. (\ref{gain}). 
\par The solution of Eq. (\ref{fp2}) will be given in Sec.~V, 
where also the mutual information transmitted by the channel will be 
evaluated. In the next section we will show 
the equivalence of the twin-beam channel with a communication scheme based 
on a couple of uncorrelated quadrature-squeezed states.
\section{Equivalence between squeezed-state and twin-beam channels}
Let us consider the unitary transformation that describes a 
50-50 frequency conversion device from mode $a$ to mode $b$. 
Under suitable choice of 
phases, the corresponding unitary 
operator $\hat U$ writes 
\begin{eqnarray}
\hat U=\exp\left[ {\pi\over 4}\left(a\bd -\ad b\right) \right] 
\;\label{fc}
\end{eqnarray}
so that the Heisenberg evolution of modes is given in matrix form as 
follows
\begin{eqnarray}
\hat U^{\dag}
\left( \begin{array}{c}a\\ b\end{array}\right)
\hat U={1\over \sqrt 2}
\left( \begin{array}{lr} 
1 & -1\\ 
1& 1\end{array}\right) 
\ \left( \begin{array}{c}a\\ b\end{array}\right)
\;\label{matrix}
\end{eqnarray}
The action of the operator (\ref{fc}) on a twin-beam state is more easily 
evaluated using Eq. (\ref{zetal}) for the twin-beam 
with $v=w=z/2$. 
Since the vacuum state $|0\>\otimes|0\>$ is an eigenvector 
of the frequency conversion operator (\ref{fc}) 
with eigenvalue 1, from the identity
\begin{eqnarray}
e^{A}\, X\, e^{-A}=X+[A,X]+{1\over 2!}[A,[A,X]]+...\; 
\end{eqnarray}
one has
\begin{eqnarray}
\hat U |z\dr _{\lambda }
&=&\left[D_a\left({i\hbox{Im}z\over \sqrt 2}\right)
S_a(\hbox{tanh}^{-1}\,\lambda) \right]|0\>\otimes
\left[D_b\left({\hbox{Re}z\over \sqrt 2}\right)
S_b(-\hbox{tanh}^{-1}\,\lambda) \right]|0\> \nonumber\\
&\equiv & |i\hbox{Im}z\>_{\lambda }
\otimes |\hbox{Re}z\>_{-\lambda }\;,\label{2sq}
\end{eqnarray}
where $S_c(r)$ is the squeezing operator for the mode $c=a,b$, namely
\begin{eqnarray}
S_c(r)=\exp\left[ {r\over 2}\left({\cd }^2-c^2\right)\right],\; 
\end{eqnarray}
and the squeezed state $|\alpha \>_{\lambda }$ 
is defined as follows
\begin{eqnarray}
|\alpha \>_{\lambda }=D_c(\alpha /\sqrt 2)\,
S_c(\hbox{tanh}^{-1}\,\lambda)|0\>\;.\label{}
\end{eqnarray}
Hence, by means of frequency conversion the twin-beam state (\ref{zetal}) 
disentangles 
into two squeezed states which are still related in intensity and 
phase. 
[For modes $a$ and $b$ at the same frequency 
and different wave vectors or polarization, this ``disentanglement'' 
can be also achieved by means of a 50-50 beam splitter \cite{paris}, 
i.e. by a passive device]. 
\par As regards the effect of frequency conversion 
at the output of a lossy/amplified 
channel, notice that a superoperator of the form
\begin{eqnarray}
{\cal L}=\alpha L[a]+\beta L[b]+\gamma L[\ad]+\delta L[\bd]+{\cal L}_{K}
\;\end{eqnarray}
undergoes the following transformation
\begin{eqnarray}
\hat U {\cal L}\hat U^{\dag}&=&{\alpha +\beta \over 2}\left(
L[a]+L[b] \right)+
{\gamma +\delta \over 2}\left(L[\ad]+L[\bd] \right) 
+{K\over 2}[a^2-{\ad}^2,\cdot]\nonumber\\
&-&{K\over 2}[b^2-{\bd}^2,\cdot] 
+{\alpha -\beta\over 2}(\hbox{cross-terms}) 
+{\gamma -\delta \over 2}(\hbox{cross-terms})
\;.\label{cross}
\end{eqnarray}
We are not interested in the cross-terms in Eq. (\ref{cross}) that 
act jointly on ${\cal H}_a\otimes{\cal H}_b$, because they vanish 
when $n_a=n_b$ and $m_a=m_b$ in Eqs. 
(\ref{loss}) and (\ref{gain}), respectively. In this case one 
has
\begin{eqnarray}
\hat U {\cal L}_{\Gamma }\hat U^{\dag}
={\cal L}_{\Gamma };\qquad \qquad\qquad \qquad
\hat U {\cal L}_{\Lambda }\hat U^{\dag}=
{\cal L}_{\Lambda }\;. 
\end{eqnarray}
Hence, the 50-50 conversion leaves the superoperators ${\cal L}_\Gamma $ 
and ${\cal L}_\Lambda $ (for both loss and PIA) invariant. 
This means that the disentanglement of the twin-beam 
occurs equivalently at whatever time during transmission. 
On the contrary, a distributed PIA with pump-correlated $a$ and $b$ 
is not invariant under the transformation (\ref{matrix}). 
Indeed, a distributed PIA followed by frequency conversion is 
equivalent to a couple of independent PSA's, 
as shown by the commutator terms in Eq. (\ref{cross}), namely 
\begin{eqnarray}
\hat U {\cal L}_{K}\hat U^{\dag}
={\cal L}'_{K}={K\over 2}[a^2-{\ad}^2,\cdot]
-{K\over 2}[b^2-{\bd}^2,\cdot] 
\;\label{pinco}
\end{eqnarray}
Notice that for $K>0$ the amplified quadrature 
components of the fields $a$ and $b$ through these commutators 
are $\hat Y_a\equiv\hat a_{\pi/2}$ and $\hat X_b\equiv
\hat b_0$, respectively, which are the right quadratures in order to 
enhance the signal carried by the couple of squeezed states (\ref{2sq}). 
Of course, the communication scheme that encodes the information on 
states (\ref{2sq})
needs two independent homodyne measurements of the quadratures 
$\hat Y_a$ and $\hat X_b$. 
\par We have shown the equivalence of the two channels as regards 
the input states and the evolution master equation. 
It remains to show that also the final detection stage is equivalent 
in the two schemes. 
The unconventional heterodyne detection is described by the following 
orthogonal resolution of the identity
\begin{eqnarray}
d\hat\mu(z,\bar z)=d^2 z\,\delta^{(2)}(\hat Z-z)\equiv 
d^2 z\,|z\dr\dl z|\;. 
\label{pomz}
\end{eqnarray}
The unitary operator (\ref{fc}) transforms 
the orthogonal resolution (\ref{pomz}) as follows
\begin{eqnarray}
\hat Ud\hat\mu(z,\bar z)\hat U^{\dag}&=&d^2 z
\,\delta(\sqrt 2\hat X_b-\hbox{Re}z)
\,\delta(\sqrt 2\hat Y_a-\hbox{Im}z)\nonumber\\
&=&dx\,dy\,\delta(\hat X_b-x)
\,\delta(\hat Y_a-y)
\;,\qquad z=\sqrt 2(x+iy)\;.
\label{pomxy} 
\end{eqnarray}
The last orthogonal resolution in Eq. (\ref{pomxy}) is just the one 
corresponding to two independent homodyne measurements of 
quadratures $\hat X_b$ and $\hat Y_a$. 
\par In conclusion of this section, we also show the equivalence 
between the QS scheme and the twin-beam scheme at the level of 
Fokker-Planck equations.
Using the homodyne probability density
\begin{eqnarray}
P_a(y;t)P_b(x;t)=\hbox{Tr}\left[\hr _t 
|y\>_{{\pi\over 2}}{}_{{\pi\over 2}}\<y|\otimes|x\>_{0}{}_{0}\<x|\right] 
\;\end{eqnarray}
the master equation
\begin{eqnarray}
{\cal L}'\hr _t&=&2\left\{A(L[\ad]+L[\bd])+
B(L[a]+L[b])\right\}\hr _t \nonumber \\
&+&{K\over 2}[a^2-{\ad}^2,\hr _t]
-{K\over 2}[b^2-{\bd}^2,\hr _t]
\;\label{me2s}
\end{eqnarray}   
can be written in the Fokker-Planck form
\begin{eqnarray}
\partial _t \left\{P_a(y;t)P_b(x;t)\right\}=
\left\{Q\left(\partial_x x+\partial_{y}y \right)
+{D\over 4}\left(\partial^{2}_{xx}+\partial^{2}_{yy}\right)\right\}
P_a(y;t)P_b(x;t)\;\label{fp2s}
\end{eqnarray}
with $Q=B-A-K$ and $D=A+B$. In obtaining Eq. (\ref{fp2s}) 
we used the Wigner rappresentation of both $a$ and $b$ modes, and then 
evaluated the marginal integration over the 
quadratures $\hat X_a$ and $\hat Y_b$. The equivalence of the Fokker-Planck 
equation (\ref{fp2s}) with Eq. (\ref{fp2}) is evident, after the 
coordinate transformation
\begin{eqnarray}
z=\sqrt 2(x+iy)\;,\qquad\qquad\qquad\qquad\bar z=\sqrt 2(x-iy)\;,\label{cha}
\end{eqnarray}
and upon renaming the product of probabilities as follows
\begin{eqnarray}
P'(z,\bar z;t)={1\over 2}P_a
\left( {\hbox{Im}z\over \sqrt 2};t\right) 
P_b\left( {\hbox{Re}z\over \sqrt 2};t\right) 
\;.\label{prod}
\end{eqnarray}
The equivalence between the two communication channels is schematized in 
Fig.~2.
\section{Evaluation of the channel capacity}
The solution of the Fokker-Planck equation (\ref{fp2})
for a Gaussian initial probability density
\begin{eqnarray}
P(z,\bar z;0)={1\over\pi\Delta^{2}(0)}
\exp{\left( -{|z-w|^2\over \Delta^{2}(0)}\right) }\;, 
\end{eqnarray}
keeps the Gaussian form at all the time, and writes
\begin{eqnarray}
P(z,\bar z;t)={1\over\pi\Delta^{2}(t)}
\exp{\left(-{|z-we^{-Qt}|^2\over \Delta^{2}(t)}\right) }\;, 
\end{eqnarray}
where the ``evolved'' variance at time $t$ is given by
\begin{eqnarray}
\Delta^{2}(t)={D\over Q}(1-e^{-2Qt})+\Delta^{2}(0) e^{-2Qt}\;. 
\end{eqnarray}
For the twin-beam communication scheme one has $\Delta ^{2}(0)\equiv
\Delta ^2_\lambda $, according to Eq. (\ref{prob}). For the scheme 
based on homodyne detection over the couple of QS states (\ref{2sq}),
one equivalently finds the Gaussian density
\begin{eqnarray}
\left|{}_0\<x|\otimes{}_{{\pi\over 2}}\<y|\hat U|w\dr _{\lambda }
\right|^2=\!{2\over \pi\Delta^2_{\lambda }}
\exp\left\{\!-
{2\over \Delta^2_{\lambda }}
\left[\left(x-{\hbox{Re} w\over \sqrt 2}\right)^2-
\left(y-{\hbox{Im} w\over \sqrt 2}\right)^2 \right]
\right\}\!\;.\label{same1}
\end{eqnarray}
\par The mutual information of a communication channel 
with conditional probability density $p(w|z)$, encoded complex 
variable $w$, and {\em a priori} distribution $p(w)$ 
writes as follows  
\begin{eqnarray}
I=\int d^2 w\int d^2 z\,p(w)p(w|z)\ln{p(w|z)\over 
\int d^2 w'\,p(w')p(w'|z)}\;. 
\end{eqnarray}
For {\em a priori} and conditional probability densities both Gaussian, i.e. 
\begin{eqnarray}
p(w)={1\over \pi\sigma ^2}\exp\left( -{|w|^2\over 
\sigma ^2}\right) \; 
\end{eqnarray}
\begin{eqnarray}
p(w|z)={1\over \pi\Delta ^2}\exp\left( -{|z-gw|^2\over 
\Delta ^2}\right) \; 
\end{eqnarray}
one has
\begin{eqnarray}
I=\ln\left( 1+{g^2\sigma ^2\over \Delta ^2}\right)
 \;. 
\end{eqnarray}
According to the Eq. (\ref{mean}) for the average number $N$ of photons 
{\em per mode}, the variance of the prior distribution for the 
twin-beam scheme under the constraint of fixed average power reads 
\cite{nota2}
\begin{eqnarray}
\sigma ^2=4\left( N-{\lambda ^2\over 1-\lambda 
^2}\right) \;. 
\end{eqnarray}
Hence, the mutual information transmitted at time $t$ for a 
twin-beam channel modeled by the Fokker-Planck equation (\ref{fp2}) is 
given by
\begin{eqnarray}
I=\ln\left( 1+
{4\left(N-{\lambda ^2\over 1-\lambda ^2}
\right)\over 
{D\over Q}(e^{2Qt}-1)+\Delta ^2 _\lambda }\right)
\;.\label{inf}
\end{eqnarray}
Eq. (\ref{inf}) is valid also for the QS scheme, with parameter 
$\lambda $ and variance $\Delta ^2_{\lambda }$ related to the squeezing 
parameter through Eqs. (\ref{2sq}) and (\ref{dellam}). 
In the lossless case, the mutual information is maximized 
by $\lambda =N/(N+1)$ to the value
\begin{eqnarray}
I=2\ln\left( 1+2N\right) \;.\label{inf2id}
\end{eqnarray}
This information is always greater than the information
of {\em two} lossless coherent-state channels, with one additional bit 
{\em per mode} for high average power ($N\gg 1$). 
For a lossy channel with damping rate $\Gamma $ and 
$n_a=n_b=\bar n$ thermal photons one has $D=\Gamma(2\bar n+1)/2$ and 
$Q=\Gamma/2$. In this case, for $\lambda =N/(N+1)$, Eq. (\ref{inf})
rewrites
\begin{eqnarray}
I=\ln\left(1+{4N\left( N+1\right)\over 
1+\left( 2N+1\right)\left(2\bar n+1\right)
\left( e^{\Gamma t}-1\right)}\right)
\;.\label{i2loss2}
\end{eqnarray}
The exponential erasure of information in Eq. (\ref{i2loss2}) 
becomes linear using the master equation (\ref{me2}), 
with $K=\Gamma /2$, namely
\begin{eqnarray}
I=\ln\left(1+{4N\left(N+1\right)\over 
1+\left( 2N+1\right)\left(2\bar n+1\right)
\Gamma\,t}\right)
\;\label{i2loss=g}
\end{eqnarray}
This case represents an ideal distributed parametric amplification 
that works against the detrimental effect of loss. 
Notice that, due to the energy of the pump, the condition of fixed 
average power is not strictly satisfied. 
Indeed, after a transient, 
the increase of the mean number of photons is approximately 
linear in time as follows
\begin{eqnarray}
N(t)={1\over 2}\left[N(0)+\bar n -{1\over 2}
+\left(N(0)-\bar n +{1\over 2} \right)e^{-2\Gamma t}
+\Gamma(2\bar n+1)\,t\right] 
\;. 
\end{eqnarray}
\section{Conclusions}
The maximum mutual information {\em per mode} for narrow-band linear 
bosonic channels under the constraint of fixed average power is 
achieved by the ideal NS channel with thermal input 
probabilities. Nevertheless, the experimental realization of number 
eigenstates and of an ideal photon number amplifier that can assist 
the channel is still 
unknown. Hence, our interest is turned into QS channels, which are 
feasible and still lead to a satisfactory efficiency. 
In this paper we have shown a new communication scheme that is based on 
unconventional-heterodyne detection on two-mode states (``twin-beam'') 
and represents an alternative way to achieve the QS channel capacity. 
We have given two experimental set-ups to 
obtain the twin-beam states. We have shown that the twin-beam 
communication channel is equivalent to a channel that employs 
a couple of QS channels and ordinary homodyne detection as decoding. 
The correspondence between the two schemes 
holds also in the presence of loss and optimal amplification. 
The equivalence is realized by a unitary transformation that 
physically corresponds to the 50-50 frequency conversion between the two 
field modes. The twin-beam scheme is easier to obtain experimentally 
as compared to the QS scheme. 
Both the encoding and decoding stages are simpler for the twin-beam 
channel than for the QS one. As encoder one just needs 
parametric downconversion and coherent states, instead of squeezing. 
On the other hand, as decoder one has just one heterodyne detector 
versus a delicate balancing of two equal homodyne detectors. 
As compared to the single homodyne QS channel, the heterodyne one, by 
employing two field modes, carries a double amount of information.

\newpage
\begin{figure}[htb]
\begin{center}
\setlength{\unitlength}{0.7cm}
   \begin{picture}(25.5,7)(-1.4,-2.5)
      \thicklines
      \drawline(-2.1,1)(0,1)
      \drawline(-2.1,2)(0,2)
      \put(-1,2.3){$a$}
      \put(-1,.5){$b$}
      \put(-1,1){\vector(1,0){0}}
      \put(-1,2){\vector(1,0){0}}
      \put(-3.5,1.8){$|0\rangle$}
      \put(-3.5,0.8){$|0\rangle$}
      \put(0,0){\framebox(4.5,3){{\footnotesize 
      $\matrix{\hbox{\large PIA}\cr
      {\hbox{gain}}\,(1-\lambda^{2})^{-1}} $}}}       
      \drawline(4.5,1)(6,1)
      \drawline(4.5,2)(6,2)
      \drawline(-1.5,-1.5)(0,0)
      \drawline(4.5,3)(6,4.5)
      \put(5.3,3.8){\vector(1,1){0}}
      \put(-0.8,-0.8){\vector(1,1){0}}
      \put(-1.5,-2){\large{pump}}
      \put(6,2.5){\large{``twin-beams''}}
      \put(7.5,1.3){\large{$|0\rangle\!\rangle_{\lambda}$}}   
      \drawline(10,1)(16.5,1)
      \drawline(10,2)(16.5,2)
      \drawline(14.1,-1)(14.1,1)
      \drawline(14.1,2)(14.1,4)
      \put(14.1,0){\vector(0,1){0}}
      \put(14.1,3){\vector(0,-1){0}}
      \drawline(13.7,0.55)(14.5,1.35)(14.4,1.45)
               (13.6,0.65)(13.7,0.55)
      \drawline(13.7,2.45)(14.5,1.65)(14.4,1.55)
               (13.6,2.35)(13.7,2.45)
      \put(13,.2){H}
      \put(13,2.5){H}
      \put(13.8,-1.9){LO}
      \put(13.8,4.4){LO}
      \put(17.4,1.3){\large{$|z\rangle\!\rangle_{\lambda}$}}   
   \end{picture}
   \begin{picture}(25.5,7)(-8.8,-2.5)
      \thicklines
      \put(-10.9,1.8){$|z/\sqrt 2\Delta_{\lambda }\rangle$}
      \put(-10.9,0.8){$|0\rangle$}
      \drawline(-9.4,1)(-6.5,1)
      \drawline(-8,2)(-6.5,2)
      \put(-8.1,1){\vector(1,0){0}}
      \put(-6.9,2){\vector(1,0){0}}
      \put(-7.3,2.3){$a$}
      \put(-7.3,.5){$b$}
      \put(-6.5,0.5){\framebox(3.5,2){{\footnotesize 
      $\matrix{\hbox{\large FC}}$}}}       
      \drawline(-7.7,-.7)(-6.5,0.5)
      \drawline(-3,2.5)(-1.8,3.7)
      \put(-2.3,3.2){\vector(1,1){0}}
      \put(-7,0){\vector(1,1){0}}
      \put(-7.7,-1.2){\large{pump}}
      \drawline(-3,1)(-1.4,1)
      \drawline(-3,2)(-1.4,2)
      \drawline(1.2,1)(3,1)
      \drawline(1.2,2)(3,2)
      \put(2.1,.5){$b$}
      \put(2.1,2.3){$a$}
      \put(2.1,1){\vector(1,0){0}}
      \put(2.1,2){\vector(1,0){0}}
      \put(-1.1,1.8){$|z/2\Delta _{\lambda}\rangle$}
      \put(-1.1,0.8){$|\bar z/2\Delta _{\lambda}\rangle$}
      \put(3,0){\framebox(4.5,3){{\footnotesize 
      $\matrix{\hbox{\large PIA}\cr
      {\hbox{gain}}\,(1-\lambda^{2})^{-1}} $}}}       
      \drawline(1.5,-1.5)(3,0)
      \drawline(7.5,3)(9,4.5)
      \put(8.3,3.8){\vector(1,1){0}}
      \put(2.3,-0.7){\vector(1,1){0}}
      \put(1.5,-2){\large{pump}}
      \drawline(7.5,1)(9,1)
      \drawline(7.5,2)(9,2)
      \put(10,1.3){\large{$|z\rangle\!\rangle_{\lambda}$}}   
      \end{picture}
\end{center}
\vskip 4truecm
\caption{Outline of the two alternative experimental set-ups to generate 
the two-mode states $|z\dr _{\lambda}$ of Eq. (\protect{\ref{sim}}). 
The labels H and LO denote high-transmissivity beam splitters and local 
oscillators respectively, which realize the displacement operators 
in Eq. (\ref{zetal}). The input states 
at the parametric amplifier (PIA) are the vacuum state in the upper 
scheme, and two independent coherent states in the lower one. 
In the last case the couple of coherent states is generated 
by a single coherent state through frequency conversion (FC).
\label{f:outline}}
\end{figure}
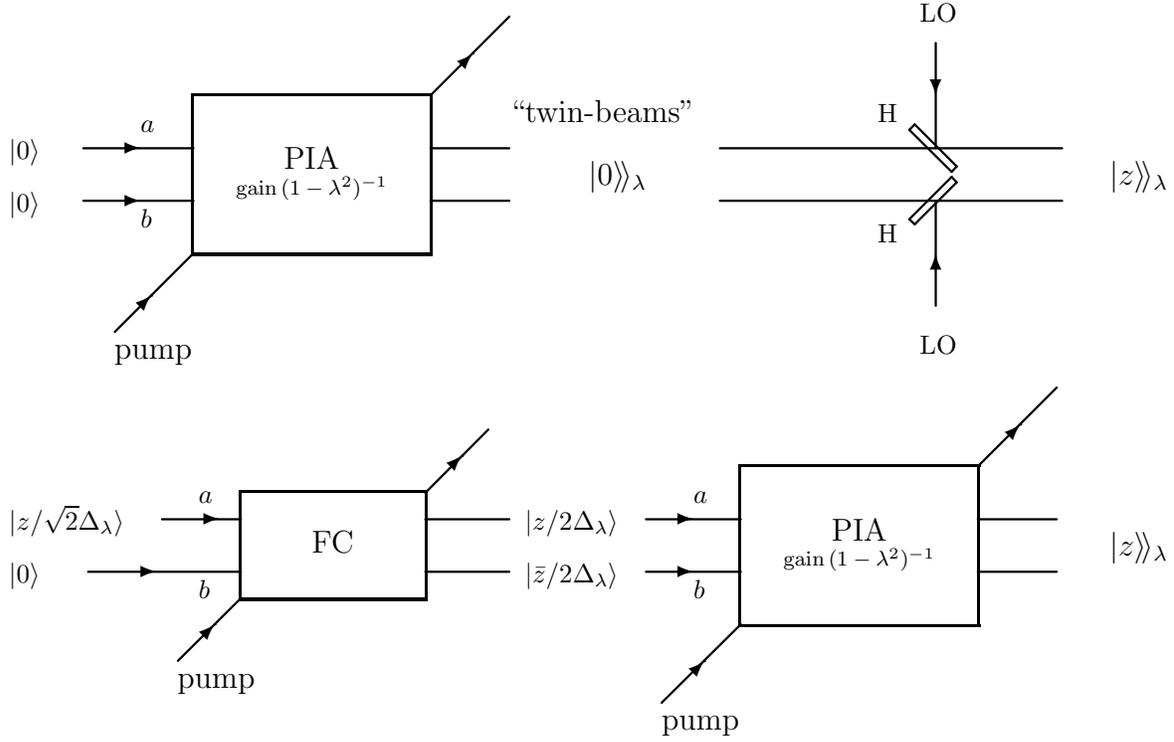
\newpage                                                                   
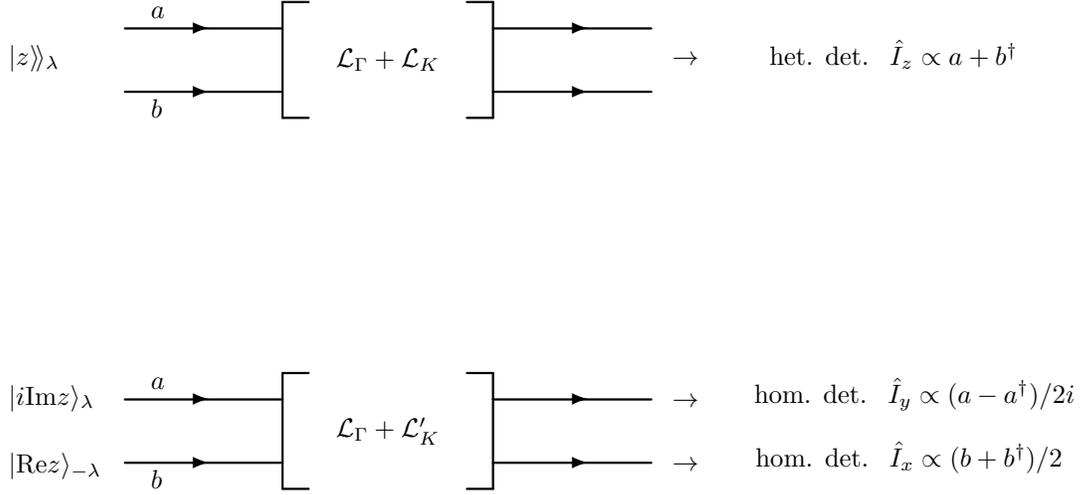
\begin{figure}[htb]
\begin{center}
\setlength{\unitlength}{0.7cm}
   \begin{picture}(25,7)(-2,-2.5)
      \thicklines
      \put(-4.2,1.4){$|z\rangle\!\rangle _{\lambda }$}
      \put(-1.5,0.4){$b$}
      \put(-1.5,2.3){$a$}
      \put(-0.4,.9){\vector(1,0){0}}
      \put(-0.4,2.1){\vector(1,0){0}}
      \drawline(-2,.9)(1,.9)
      \drawline(-2,2.1)(1,2.1)
      \put(3,1.5){\makebox(0,0){${\cal L}_{\Gamma}+{\cal L}_{K}$}}
      \drawline(1.5,.4)(1,.4)(1,2.6)(1.5,2.6)
      \drawline(4.5,.4)(5,.4)(5,2.6)(4.5,2.6)
      \drawline(5,.9)(8,.9)
      \drawline(5,2.1)(8,2.1)
      \put(6.8,.9){\vector(1,0){0}}
      \put(6.8,2.1){\vector(1,0){0}}
      \put(8.4,1.4){$\rightarrow $}   
      \put(12.6,1.6){\makebox(0,0){\hbox{het. det. }
         ${\hat I}_z \propto a+b^{\dag}$}}
   \end{picture}
   \begin{picture}(25,7)(-2,-2.5)
      \thicklines
      \put(-4.2,2){$|i\hbox{Im}z\rangle _{\lambda }$}
      \put(-4.2,.7){$|\hbox{Re}z\rangle _{-\lambda }$}
      \put(-1.5,0.4){$b$}
      \put(-1.5,2.3){$a$}
      \put(-0.4,.9){\vector(1,0){0}}
      \put(-0.4,2.1){\vector(1,0){0}}
      \drawline(-2,.9)(1,.9)
      \drawline(-2,2.1)(1,2.1)
      \put(3,1.5){\makebox(0,0){${\cal L}_{\Gamma}+{\cal L}'_{K}$}}
      \drawline(1.5,.4)(1,.4)(1,2.6)(1.5,2.6)
      \drawline(4.5,.4)(5,.4)(5,2.6)(4.5,2.6)
      \drawline(5,.9)(8,.9)
      \drawline(5,2.1)(8,2.1)
      \put(6.8,.9){\vector(1,0){0}}
      \put(6.8,2.1){\vector(1,0){0}}
      \put(8.4,.75){$\rightarrow $}   
      \put(8.4,1.95){$\rightarrow $}   
      \put(13,2.2){\makebox(0,0){\hbox{hom. det. }
         ${\hat I}_y \propto (a-a^{\dag})/2i$}}
      \put(12.9,1){\makebox(0,0){\hbox{hom. det. }
         ${\hat I}_x \propto (b+b^{\dag})/2$}}
      \end{picture}
\end{center}
\vskip 4truecm
\caption{Equivalence between the channels transmitting twin-beam states 
$|z\dr _{\lambda}$ and couples of quadrature-squeezed states 
$|i\hbox{Im}z\rangle _{\lambda }\otimes 
|\hbox{Re}z\rangle _{-\lambda }$. 
The decoding measurements are the unconventional-het\-ero\-dyne detection 
(complex photocurrent $\hat I_z$) 
and two independent ordinary homodyne detections 
(real photocurrents $\hat I_x$ and $\hat I_y$), respectively. 
The superoperators ${\cal L}_{\Gamma},\ {\cal L}_{K}$ and ${\cal L}'_{K}$ 
model the loss in Eq. (\ref{loss}), the phase-insensitive 
amplification in Eq. (\ref{comm}) and the phase-sensitive 
amplification in Eq. (\ref{pinco}), respectively. 
The equivalence between the two communication schemes is realized 
by the unitary transformation (\protect{\ref{matrix}}), 
namely a 50-50 frequency conversion.\label{f:equiv}}
\end{figure}                                                                   

\newpage
\begin{thebibliography}{99}
\bibitem{hol} A. S. Holevo, Probl. Inf. Trans. {\bf 9}, 177 (1973).
\bibitem{yoz} H. P. Yuen and M. Ozawa, Phys. Rew. Lett. {\bf 70}, 363 (1993).
\bibitem{caves} C. M. Caves and P. D. Drummond, Rew. Mod. Phys. {\bf 66}, 
481 (1994).
\bibitem{yy} H. P. Yuen, Phys. Rew. Lett. {\bf 56}, 2176 (1986).
\bibitem{gmd} G. M. D'Ariano, Phys. Rev. A {\bf 45}, 3224 (1992).
\bibitem{nota} The NS channel in the presence of loss 
has been optimized in Ref. \cite{next}. 
In Ref. \cite{hall} Hall has analyzed the degradation 
due to additive Gaussian noise (which models general environmental 
effects due to linear interactions with random fields).
\bibitem{next} G. M. D'Ariano and M. F. Sacchi, unpublished. 
\bibitem{hall} M. J. W. Hall, Phys. Rew. A {\bf 50}, 3295 (1994).
\bibitem{shap} J. H. Shapiro, Opt. Lett {\bf 20}, 1059 (1995). 
\bibitem{yush} H. P. Yuen and J. H. Shapiro, IEEE Trans. Inform. Theory 
IT {\bf 26}, 78 (1980).
\bibitem{shwa} J. H. Shapiro and S. S. Wagner, IEEE J. Quantum Electron. 
QE {\bf 20}, 803 (1984).
\bibitem{rc} G. M. D'Ariano and M. F. Sacchi, Phys. Rev. A {\bf 52}, 
R4309 (1995).
\bibitem{mauro1} G. M. D'Ariano, Int. J. Mod. Phys. B {\bf 6}, 1292 (1992).
\bibitem{rc2} Indeed in Ref. \cite{rc} the twin-beam state is displaced 
asymmetrically only with respect to one mode. As a result of the 
symmetric displacement here we gain an additional factor $1/\sqrt 2$
in phase sensitivity.
\bibitem{proc} G. M. D'Ariano and M. F. Sacchi, 
{\em Repeatable two-mode phase measurement} in {\em Quantum 
Interferometry}, Proceedings of an Adriatico Workshop, Trieste, March 
1996, ed. by F. De Martini, G. Denardo, Y. Shih, Verlagsleiter 
(Weinheim 1996), pp. 307-313.
\bibitem{lind} G. Lindblad, Commun. Math. Phys. {\bf 48}, 119 (1976).
\bibitem{gard} C. W. Gardiner, {\em Quantum noise}, Springer-Verlag, 
Berlin Heidelberg, 1991.
\bibitem{paris} M. G. A. Paris, Phys. Lett. A {\bf 225}, 28 (1997).
\bibitem{nota2} By allowing the gain parameter $\lambda $ to be varied 
as a function of the ``displacement signal'' $z$, one might achieved a 
little better capacity. This is not easy to evaluate because the 
conditional probability would not have a Gaussian form anymore. 
Moreover, such a varying parameter would be very difficult to 
get experimentally.
\end{thebibliography}
\end{document}